\documentstyle[astrobib]{mnv2}
 
\input epsf
\input rotate
\epsfverbosetrue

\newcommand{\ltapprox}{\raisebox{-0.5ex}{$\,\stackrel{<}{\scriptstyle \sim}\,$}}
\newcommand{\gtapprox}{\raisebox{-0.5ex}{$\,\stackrel{>}{\scriptstyle \sim}\,$}}

\title[Dense, thin clouds in AGN]
      {Dense, thin clouds and reprocessed radiation
       in the central regions of Active Galactic Nuclei}
\author[Kuncic, Celotti and Rees] 
       {Z. Kuncic$^{1}$\thanks{Present address: ANU Astrophysical Theory
        Centre, School of Mathematical Sciences, Australian National
        University, ACT 0200, Australia,
        E-mail: {\tt kuncic@maths.anu.edu.au}},
        A. Celotti$^{1,2}$ and M. J. Rees$^{1}$ \\ 
        $^{1}$Institute of Astronomy, Madingley Rd, Cambridge CB3 0HA \\
        $^{2}$SISSA/ISAS, via Beirut 4, 34014 Trieste, Italy }
\date{}
\pubyear{1996}

\begin{document}

\maketitle

\begin{abstract}
The primary radiation generated in the central continuum-forming region
of Active Galactic Nuclei (AGN) can be reprocessed by very dense,
small-scale clouds that are optically-thin to Thomson scattering.
In spite of the extreme conditions expected to prevail in this innermost,
central environment, the radiative clouds can survive and maintain cool
temperatures relative to the ambient emitting region by means of
confinement by strong magnetic fields.
Motivated by these ideas, we present a detailed quantitative study of
such clouds, explicitly describing the physical properties they can
attain under thermal and radiative equilibrium conditions.
We also discuss the thermal stability of the gas in comparison to that of
other reprocessing material thought to reside at larger distances from the
central source.
We construct a model to predict the emergent spectra from a source
region containing a total line of sight column density
$\la 10^{23} {\rm cm}^{-2}$ of dense clouds which absorb and
reemit the primary radiation generated therein.
Our predicted spectra show the following two important results:
(i) the reprocessed flux emitted at optical/UV energies is insufficient
to account for the blue bump component in the observed spectra; and
(ii) the amount of line radiation that is emitted is at least comparable
to (and in many cases dominates) the continuum radiation.
The lines are extremely broad and tend to accumulate in the extreme
ultraviolet (EUV), where they form a peak much more prominent
than that which is observed in the optical/UV.
This result is supported by current observations, which indicate
that the spectral energy distribution (SED) of radio-quiet AGN may
indeed reach a maximum in the EUV band.
\end{abstract}


\section{Introduction}

There is mounting observational evidence that the centres of most AGN
harbour vast quantities of thermal matter and that this material plays
a key role in the generation of the observed spectra by reprocessing
the primary radiation.
The importance of reprocessing by relatively cool ($\sim 10^{5}$ K),
dense gas in the central environments of AGN is clearly evident from
the observed X-ray spectra, particularly those of Seyfert nuclei, which
typically exhibit a variety of distinct thermal signatures in the form of
lines, edges and reflection features
(e.g. \citeNP{Pounds90,Nandra91,NP94}).
However, the most compelling evidence for the presence of
energetically-significant amounts of cool material in the central
regions of radio-quiet quasars and Seyferts (which comprise the vast
majority of AGN) is provided by the excess of optical/UV over X-ray
continuum radiation known as the `blue bump'
\cite{Shields78,EdelMalk86,Elvis94}.
This is typically the most prominent component in the SED of these
objects and can contain as much as half of the entire observed spectral
power (e.g. \citeNP{Neugebauer87,Sanders89,Kolman93}).
The soft X-ray excess that is often seen above an extrapolation of the
X-ray power law \cite{Arnaud85,WilkesElvis87,WalterFink93} may be the
high energy tail of the blue bump and indeed, recent observations
imply a link between the two components, with the possibility that they
constitute a single `big bump' component extending from the optical/UV
band up to soft X-rays (see \citeNP{Puch96} and references cited therein).

In the framework of the standard black hole paradigm for AGN
\cite{Rees84}, any optically-thick thermal gas residing within about 100
gravitational radii, regardless of its distribution, is expected to produce
a quasi-blackbody spectrum that peaks at $\sim 10^{5}$ K.
Since the bulk of the accreting material is expected to fall into an
optically-thick disk, this is the most natural explanation for the blue
bump \cite{Shields78,MalkanSargent82,CzernyElvis87,SunMalkan89}.
Although the thermal character of the blue bump is now independently
confirmed \cite{Koratkar95}, detections of quasi-simultaneous optical
and UV variations (e.g. \citeNP{CC91,Clavel92}) are inconsistent with
the radiation being `intrinsic' to a standard accretion disk (i.e. emitted
as a result of internal dissipation).
These variability measurements suggest instead that the blue bump is
probably `secondary' radiation that has been reprocessed by the disk
(e.g. \citeNP{LightmanWhite88,GeorgeFabian91}).
Indeed, the presence of relatively cool reprocessing material in a disk
within 20 gravitational radii of a central mass is inferred from the very
recent ASCA X-ray observations \cite{Tanaka95}.

However, not all cool gas at the centres of AGN need necessarily reside
in the putative accretion disk.
Some may be distributed in the form of numerous dense clouds or filaments
which occupy only a small fraction of the total volume of the source region
but which provide a substantial covering factor \cite{GR88}.
Indeed, several models which calculate the reprocessed spectra due to
optically-thick clouds are successful in reproducing the general
features of the observed X-ray spectra
(e.g. \citeNP{SivTsu93,BondMatsuoka93,NandraGeorge94}).
However, the absence of a strong soft X-ray cutoff in the observed spectra
of some objects limits the line of sight column densities to
$\la 10^{21} {\rm cm}^{-2}$ (e.g. \citeNP{TurnerPounds89}) or to
slightly higher values ($\sim 10^{23} {\rm cm}^{-2}$) if the gas is warmer
and hence, more highly-ionized \cite{Nandra91}.
This means that clouds with a significant Thomson optical depth residing
within the primary source region where the X-rays are produced must
have a covering factor less than unity so that not all the X-rays are
intercepted, as implied by the soft X-ray absorption limits.

Clouds can be distributed throughout the central source region with a
large covering factor (close to unity) if they are optically-thin to
Thomson scattering.
These clouds might contribute to the optical/UV blue bump
(\citeNP{CFR92}, hereafter CFR92) and indeed, some models even attribute
the blue bump to such reprocessing material (\citeNP{Barvainis93}).
The presence of Thomson-thin reprocessing gas could also be postulated
from the observed spectra of quasars, which typically lack a reflection
component, but which exhibit a blue bump that is energetically more
significant than the X-ray flux \cite{Williams92,Elvis94}.
However, Thomson-thin clouds must be extremely dense in order to keep
cool and to produce distinct thermal signatures in the observed spectra.
As examined by Ferland \& Rees (1988; hereafter FR88), the radiative
equilibrium of very dense gas irradiated by an intense AGN source is
dominated by free-free absorption.
Induced processes can also be important, as well as photoionization and
recombination, which are the dominant processes for the cooler gas in
the more distant broad line region (BLR) in AGN
(see e.g. \citeNP{Osterbrock85}).

In the central continuum-forming region of AGN, a plausible scenario
is that dense clouds are embedded within a hot, magnetically-dominated
gas which pervades the source region, forming a corona above and below
the inner disk (e.g. \citeNP{SvZd95,ZdzMag96}).
The strong magnetic fields which could be responsible for accelerating
the radiating particles can also provide the chief means of cloud
confinement required in such a multi-phase medium \cite{Rees87}.
Moreover, the confining magnetic stresses can, by suppressing transverse
conductivity, effectively insulate the clouds from the external hot plasma
and, in the presence of strong radiation forces, can maintain the very small
scalesizes that are implied by the high densities and the column density
constraint inferred from the lack of strong soft X-ray extinction.
In this case `clouds' may ultimately manifest themselves as magnetospheric
filaments or streaks, depending on the field configuration (CFR92; see 
also \citeNP{KBR96}).

Indeed, the presence of dense clouds in sources where the magnetic and
radiation fields are in equipartition can explain the lack of correlated
variability between IR/optical and X-ray fluxes (detected in at least one
object -- see \citeNP{CGF91}), since free-free absorption is capable of
depleting all of the low frequency radiation from a primary source.
When the free-free absorption extends up to optical/UV frequencies
($\sim 10^{15}$ Hz), it is energetically possible that reprocessing by
free-free absorbing clouds can ultimately produce a significant fraction
of the blue bump, as suggested by CFR92.

At temperatures $\sim \! 10^{5}$ K, the radiation reemitted from such gas
is expected to comprise free-free (bremsstrahlung) and bound-free
emission due to hydrogen and, to a lesser extent, helium nuclei, as well
as line emission (bound-bound transitions) due to heavier nuclei.
In practice, it is necessary to consider the thermal and ionization balance
of the reprocessing gas as well as radiative transfer effects in order to
determine the relative importance of these processes and to predict the
emergent spectra from a collection of clouds.

In this paper, we use a numerical code to obtain explicitly the physical
properties (temperature and ionization structure) acquired by very dense
clouds in thermal and radiative equilibrium and to calculate the
emergent spectra from a source region containing such clouds.
Our objective is to further study and extend the ideas presented by CFR92
by determining whether cool temperatures and small column densities are
compatible radiative equilibrium properties of dense clouds which might
reside in an AGN magnetosphere.
We reexamine an earlier attempt by FR88 at predicting the spectra from
similar clouds by constructing a more self-consistent model which
explicitly takes into account line emission and the reprocessed radiation
field.
We also compare in closer detail the predicted spectra with current
observations.
In Section 2, we state the specifications and assumptions of our
numerical analysis, while in Sections 3 and 4, respectively, we study
the thermal structure and stability of isolated dense clouds.
In Section 5, the emergent spectra predicted by our cloud model are
presented and compared with the observed SEDs of radio-quiet AGN.
We summarize our results and their implications in Section 6.

\section{Assumptions and Specifications}

\subsection{Radiative Equilibrium}

Dense clouds residing at the centres of AGN can attain radiative
equilibrium and thereby contribute to the observed spectrum by absorbing
and reemitting the primary radiation provided that they can survive for
at least a few radiative lifetimes, $t_{\rm rad}$.
This is the case for a wide range of reasonable parameters.
The clouds must be small enough to come into pressure equilibrium with
their surroundings, which requires the sound-crossing time to be shorter
than the characteristic dynamical timescale, $t_{\rm dyn}$.
On the other hand, the clouds cannot be so small that they are too quickly
destroyed by microphysical diffusion processes \cite{KBR96}.
For the clouds to sustain cool temperatures, there is also the requirement
that $t_{\rm rad}$ must be shorter than $t_{\rm dyn}$.
The longest radiative cooling timescale for partially-ionized gas with
an electron temperature $T_{\rm e}$ and number density $n_{\rm e}$ is
typically that for bremsstrahlung emission due to electron encounters
with ions,
\[
t_{\rm rad} \sim 10^{13} n_{\rm e}^{-1}
    \left( \frac{T_{\rm e}}{10^{5}{\rm K}} \right)^{1/2}
    \, \, {\rm s} \, .
\]
An indication on $t_{\rm dyn}$ in the primary source region comes from
the observed timescale for X-ray variability.
This is typically on the order of hours, although variability timescales
as short as 100 s have been measured in some objects
(see \citeNP{MDP93} for a review).
Thus, over these timescales, thermal gas with electron densities
$n_{\rm e} \gg 10^{11} {\rm cm}^{-3}$ can maintain radiative equilibrium
in the central regions of AGN.

Since the absorption and emission rates of the specific radiation
processes have a complex temperature dependence and the total heating
and cooling rates further depend on the coupling between the local
temperature and ambient radiation field, as described by the radiative
transfer equation, a self-consistent thermal and radiative equilibrium
solution $T_{\rm e}$ can only be evaluated with the use of statistical
equations and numerical iterative methods.
We employ the code {\small CLOUDY} (version 84.09) \cite{Ferland93},
which calculates the radiative equilibrium properties of thermal gas
that is optically-thin to Thomson scattering.
This code has been previously tested by FR88 for its application to the
case of an intense nonthermal AGN source irradiating clouds with a range
of densities.
The code calculates a self-consistent equilibrium electron temperature
$T_{\rm e}$ at discrete points within a cloud of given density and size,
for certain specifications and assumptions about the total flux and
spectral shape of the incident radiation, the chemical composition of
the gas and the  cloud geometry.
These are described next.

\subsection{The Primary Radiation Field}

The high-energy spectra of radio-quiet quasars and Seyferts are
understood to comprise a reprocessed component that is flatter
in the 2-30 keV range (e.g. \citeNP{Pounds90,NP94}).
Taking this into account, we assume a primary radiation field composed
of a power-law component which extends over X-ray energies with a spectral
index $\alpha \equiv - {\rm d} \log F_{\nu} / {\rm d} \log \nu = 1.0$
that is consistent with the observed spectra when the reprocessing
features are decoupled.
We assume that this primary spectrum steepens to an index of $2.0$
above a break energy in hard X-rays at 100 keV, as predicted by the
currently-favoured thermal Comptonization models
(e.g. \cite{HaardtMaraschi93,Stern95}).
Although there is some evidence from typical Seyfert 1 spectra for such
a high-energy cutoff, it is not very well constrained by the observations
(e.g. \citeNP{Maisack93,Johnson94,Zdz95}).
However, the uncertainty in its precise location will not affect the
radiative equilibrium of the dense gas we investigate here, since with
the low column densities we consider, the gas is only weakly coupled to
such hard radiation.
At low frequencies, we assume a self-absorbed synchrotron spectrum with
a FIR turnover frequency $\nu_{\rm t}$ (see \citeNP{deKoolBegel95}).

Specifically, we describe this primary radiation with a mean intensity
(${\rm erg.s}^{-1}{\rm cm}^{-2}{\rm Hz}^{-1}$)
\[
j_{\nu} \propto
	\left\{
	\begin{array}{llc}
	\nu^{5/2} & , & h\nu \ltapprox h\nu_{\rm t} \\
  \nu^{-1} & , & h\nu_{\rm t} \ltapprox h\nu \ltapprox 100 \mbox{ keV} \\
	\nu^{-2} & , & h\nu \gtapprox 100 \mbox{ keV} \, .
	\end{array}
	\right.
\]
In the present context, it is instructive to normalize this primary
spectrum with a total flux expressed in terms of a radiation energy-density
temperature $T_{\rm rad}$, since this is just the equivalent blackbody
temperature which is attained in the optically-thick limit, defined by
(see FR88)
\[
a T_{\rm rad}^{4} = \frac{4 \pi}{c} \int d\nu \, j_{\nu}
            = \frac{L}{4 \pi r^{2} c} \, ,
\]
where $L$ is the total bolometric luminosity, $r$ is the size of the
source region and $a$ is the usual radiation energy-density constant.
To explore the properties of clouds in high and low radiation energy-density
environments, we use the limits of the range
$5 \times 10^{4}$ K $\ltapprox T_{\rm rad} \ltapprox 2 \times 10^{5}$ K.
The upper and lower bounds of this range would correspond to a source of
size $r \sim 10r_{\rm g}$ and $r \sim 100r_{\rm g}$, respectively,
radiating at the Eddington limit of a black hole of mass $10^{7}M_{\odot}$,
with a gravitational radius $r_{\rm g} \sim 1.5 \times 10^{12}$ cm.

Because of the high intensity of the primary radiation field in the
central continuum-forming region, it is of interest to examine the
effects of different maximum brightness temperatures,
$T_{\nu} \equiv (c^{2}/2k\nu^{2})j_{\nu}$, and to do so,
we choose two extreme values of the self-absorption turnover frequency:
$\nu_{\rm t} = 3 \times 10^{13}$ Hz and $\nu_{\rm t} = 5 \times 10^{14}$ Hz.
The value of $\nu_{\rm t}$ affects gas heating, not only via free-free
absorption, but also via Compton scattering, which can be enhanced by the
extra contribution of induced scattering.
Induced Compton scattering has an optical depth which differs from the
Thomson optical depth $\tau_{\rm T}$ by a factor $kT_{\nu}/m_{\rm e}c^{2}$;
it can be important relative to ordinary (spontaneous) Compton scattering
when $\tau_{\rm T} < 1$ \cite{LevSun70} and provides a heat input (rather
than cooling) even when $h\nu \ll kT_{\rm e}$.
A high-$T_{\nu}$ radiation field can induce Compton scattering of the
photons in the peak of the $T_{\nu}$ distribution, shifting them to lower
frequencies \cite{Sunyaev71}.

\begin{figure*}
\hspace{0.001\textwidth}
\epsfysize=0.95\textwidth
\setbox1=\vbox{\epsfbox{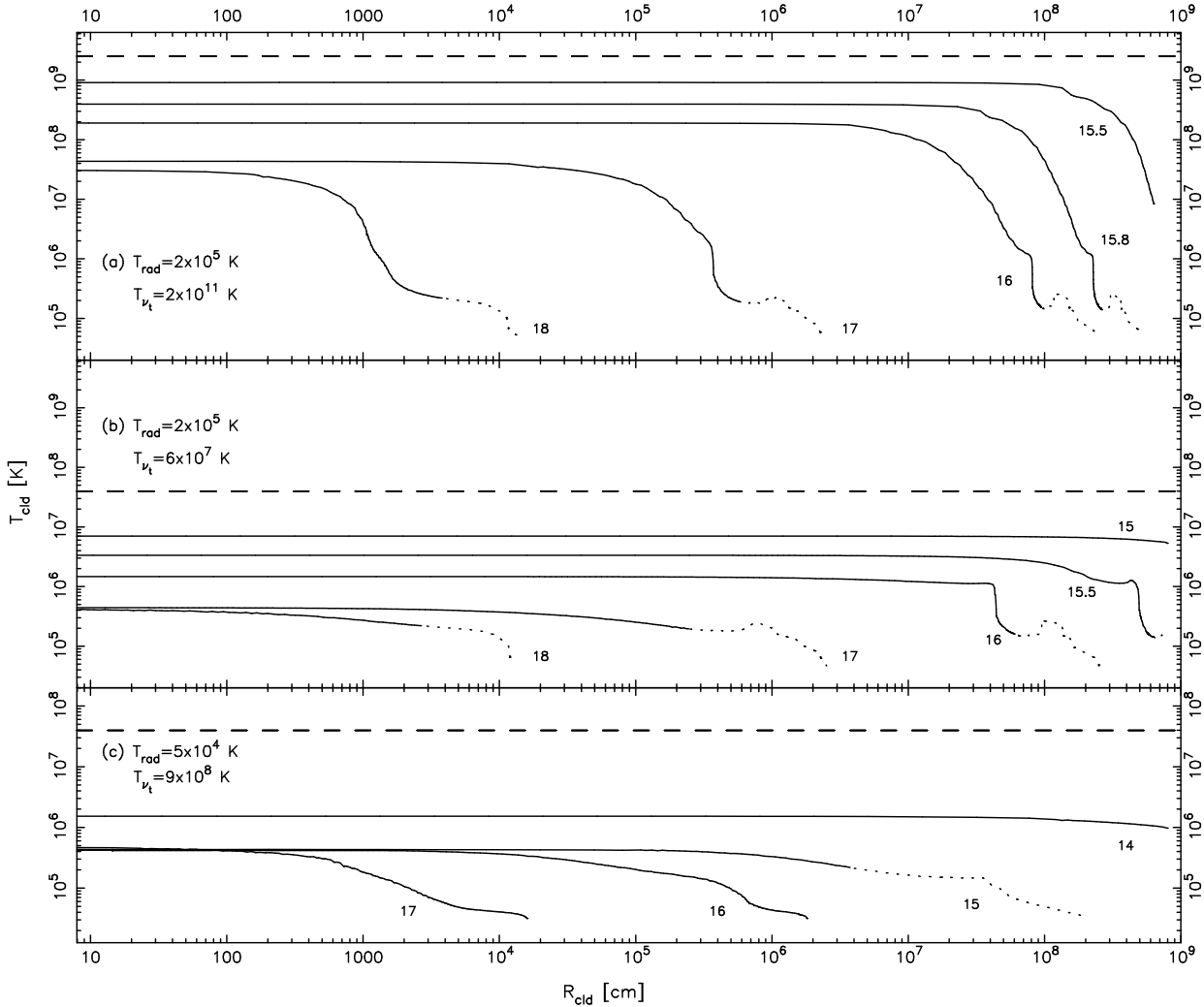}}
\rotr1
\caption{Cloud temperature $T_{\rm cld}$ plotted as a function of cloud
depth $R_{\rm cld}$ at constant hydrogen density ($\log n_{\rm H}$
indicated on each curve) for two limiting values of the radiation
energy-density temperature $T_{\rm rad}$ and for different maximum
radiation brightness temperatures $T_{\nu_{\rm t}}$.
The dashed line in each plot indicates the corresponding Compton
temperature $T_{\rm C}$.
The dotted regions along the curves indicate
where the cloud has an optical depth to free-free absorption that is
greater than unity at $10^{15}$ Hz. }
\label{fig:profiles}
\end{figure*}

\subsection{Cloud Assumptions}

Assuming a standard solar composition (\citeNP{GrevesseAnders89} is
used by the code), we fix the total hydrogen density of the cloud gas
at constant values $n_{\rm H} \gtapprox 10^{15} \mbox{ cm}^{-3}$ for
which radiative processes dominate Compton scattering \cite{FerlandRees88}.
We choose the condition of constant density rather than constant pressure
because, as we explain below in Section \ref{sec:stability}, magnetic
confinement of clouds allows a range of possible thermal pressures.
The code reliably handles densities up to $10^{18} \mbox{ cm}^{-3}$,
although then only at temperatures which are not too far below $10^{5}$ K.
Since individual clouds are always several orders of magnitude
smaller than the scalesize $r$ of the source, a plane-parallel cloud
geometry is an effective working approximation.
The code then calculates the equilibrium electron temperature at discrete
points separated by a thickness sufficiently small for the physical
conditions within to be essentially uniform.
The resulting cloud temperature $T_{\rm cld}$ is taken to be the equilibium
electron temperature in the final of these zones, at the far side
(leading edge) of the cloud.
Since the relative importance of all the radiative processes operating in
a cloud depends upon the optical depth, the physical conditions vary from
one point to the next.
It is therefore necessary to examine the thermal structure of the clouds
in order to identify the dominant processes as a function of depth
$R_{\rm cld}$ into the gas.

\section{The Thermal Structure of a Cloud}

Fig. \ref{fig:profiles} shows log-log plots of $T_{\rm cld}$ as a function
of $R_{\rm cld}$ for an isolated cloud of gas subjected to the primary
radiation, at the two limiting values of $T_{\rm rad}$.
The profile curves have been calculated up to a value of $R_{\rm cld}$
which either corresponds to a column density of at least
$10^{23} {\rm cm}^{-2}$ or is the maximum value allowed by the code for
the specified density and radiation field.
The dashed line in each plot indicates the Compton temperature, $T_{\rm C}$,
of the primary radiation field.
In Fig. \ref{fig:profiles}$a$, this is the initial $T_{\rm C}$, since
induced Compton scattering is important in this case, as we describe
below.
The dotted regions on the profiles indicate where the optical depth to
free-free absorption at a frequency $10^{15}$ Hz is greater than unity.
For convenience, we use this criterion to distinguish between an
`optically-thin' and an `optically-thick' cloud in the following.
The most immediate result to be noted from these profiles is that they
show there exist thermal and radiative equilibrium solutions for clouds
of cool gas with very small scalesizes.

\subsection{High Energy-Density Environments}

The temperature profiles of an isolated cloud at various constant densities
$n_{\rm H}$ and in a high $T_{\rm rad}$ environment are shown in
Figs. \ref{fig:profiles}$a$ and \ref{fig:profiles}$b$.
These figures differ in the value of the self-absorption turnover frequency
specified in the primary radiation field, with
$\nu_{\rm t} = 3 \times 10^{13}$ Hz in Fig. \ref{fig:profiles}$a$ and with
$\nu_{\rm t} = 5 \times 10^{14}$ Hz in Fig. \ref{fig:profiles}$b$.
These values correspond to radiation fields with a maximum brightness
temperature $T_{\nu_{\rm t}} = 2 \times 10^{11}$ K and
$T_{\nu_{\rm t}} = 6 \times 10^{7}$ K, respectively.
In Fig. \ref{fig:profiles}$a$, the initial Compton temperature of the
radiation field is $T_{\rm C} = 3 \times 10^{9}$ K, which is considerably
higher than $T_{\rm C} = 7 \times 10^{7}$ K for Fig. \ref{fig:profiles}$b$
owing to the additional contribution to Compton heating by induced
scattering, which completely overwhelms ordinary (spontaneous) scattering
in the high-$T_{\nu_{\rm t}}$ case, but is negligible in the
low-$T_{\nu_{\rm t}}$ case.

\subsubsection{Optically-Thin Cloud}

At densities $\ltapprox 10^{16} \mbox{ cm}^{-3}$, the temperature of
the gas at its illuminated edge approaches the Compton temperature. 
This is because as the density falls, collisional processes
(free-free absorption, bremsstrahlung emission and collisionally-excited
line emission) become increasingly less efficient and electron scattering
makes an increasing contribution to the heating of the gas.
Although free-free absorption accounts for essentially all of the
heating at the highest densities and smallest $R_{\rm cld}$, a comparison
of the curves for $\log n_{\rm H} = 18$ in Figs. \ref{fig:profiles}$a$
and \ref{fig:profiles}$b$ shows that induced scattering is still capable
of giving rise to a very high temperature (a few $\times 10^{7}$ K) at
the illuminated edge.
At such high densities, however, induced scattering can only maintain
the high temperatures over a short depth into the gas structure, since
the free-free absorption rapidly depletes the energy in the FIR peak of
the primary spectrum.

As the density falls below  $10^{16} \mbox{ cm}^{-3}$ and free-free
absorption becomes negliglible, the gas can remain at $> 10^{8}$ K
out to a column density where the optical depth to induced scattering
at the FIR peak is of order unity.
Beyond this depth, which is typically
$(\mbox{a few } \times 10^{6}) n_{16}^{-1}$ cm for the parameters in
Fig. \ref{fig:profiles}$a$
(where $n_{16} = n_{\rm H}/10^{16}{\rm cm}^{-3}$),
induced scattering begins to redistribute photons from the FIR peak.
Consequently, the electron temperature gradually falls and approaches
the value it would attain in the absence of induced scattering, as shown
by the corresponding curves in Fig. \ref{fig:profiles}$b$.

The cooling at densities $\gg 10^{16}{\rm cm}^{-3}$ is chiefly regulated
by collisional processes.
At temperatures $> 10^{7}$ K, the gas is very nearly completely ionized
and consequently, the primary cooling occurs via free-free transitions
(bremsstrahlung emission due to H and, to a lesser extent, He nuclei),
giving rise to continuum X-ray radiation.
Line radiation contributes typically $\ltapprox 30\%$ to the total
radiative cooling in gas at these extreme temperatures.
This radiation is chiefly comprised of atomic resonance lines arising
from collisionally-excited Fe nuclei at ionization states XVI and above
and re-emitted in the EUV band.
As the temperature falls below $10^{7}$ K, the lines are emitted by
successively lower ionization states of Fe and the efficiency of
bremsstrahlung decreases, although it remains the dominant coolant
until $T_{\rm cld}$ falls below $10^{6}$ K.

Gas with an initial temperature $> 10^{6}$ K at its illuminated edge
eventually cools to this temperature.
There is a distinct break in the temperature profiles at the
corresponding column density, indicating where the ionization level
can no longer be maintained (see Fig. \ref{fig:profiles}$a$ and
\ref{fig:profiles}$b$ at $n_{\rm H} < 10^{17}{\rm cm}^{-3}$) .
The lines which continue to cool the gas are emitted mainly by Fe in
much lower ionization states (down to FeIX at 171 \AA).
These lines can account for up to $50\%$ of the total radiative cooling
in an optically-thin structure, dominating bremsstrahlung emission once
the temperature falls below $10^{6}$ K.
The ionization front is most evident for $\log n_{\rm H} \ltapprox 16$,
since the column densities attained before the gas appreciably cools
are sufficiently large to trap the small cross-section, high-energy photons
(at soft X-rays) that ionize the heavy element constituents.
Photoabsorption of soft X-rays is an important source of heating for the
gas as it passes through the ionization front.
When the gas temperature at the illuminated edge is $< 10^{6}$ K and
hence, there is no ionization front
(see Fig. \ref{fig:profiles}$b$ for $\log n_{\rm H} \gtapprox 17$),
photoabsorption is important throughout the entire gas structure and
provides an increasing contribution to the total heating, reaching up to
$40\%$ before the cloud becomes optically-thick.

At densities $\gtapprox 10^{17} {\rm cm}^{-3}$, radiative cooling is
efficient enough to cool the gas to temperatures approaching $10^{5}$ K
before the structure becomes optically-thick.
The cooling processes which become effective at these temperatures are
collisionally-excited Ly $\alpha$ line emission by HeII (304 \AA),
induced HII recombination and, at $\log n_{\rm H} = 18$, MgVII (433 \AA)
line emission.
In addition to H free-free absorption and photoabsorption of metals, about
$10\%$ of the total heating is due to collisional ionization of H from
excited levels.

\subsubsection{Optically-Thick Cloud}

As evident in Figs. \ref{fig:profiles}$a$ and \ref{fig:profiles}$b$, dense
gas in a high $T_{\rm rad}$ environment attains an optically-thick structure
at temperatures typically just above $10^{5}$ K.
At the highest densities, the heating is entirely dominated by free-free
absorption.
Photoabsorption becomes significant for
$\log n_{\rm H} \ltapprox 17$ and gives rise to a distinctive peak in the
corresponding temperature profiles.
This peak does not emerge at the highest densities because line emission
resulting from collisional excitations is efficient enough to cool the gas
before it attains column densities sufficiently large for photoabsorption
to become significant.
The dominant lines are due to HeII (Ly $\alpha$, 304 \AA), FeIX (171 \AA)
and MgVII (433 \AA).

The peak is followed by a drop in the temperature towards $10^{5}$ K,
which is close to the ionization potentials of HII and HeII.
At the corresponding depths, these nuclei contribute
to the cooling of the gas through bremsstrahlung emission, recombination
(including induced recombination of HII), HeII collisional ionization and
Ly $\alpha$ emission.
Collisional ionization of heavier elements (mostly C, N, O) also
contributes to the cooling.
Although these processes regulate essentially all of the cooling of
optically-thick gas when $\log n_{\rm H} \ltapprox 17$, they only account
for up to about $30\%$ of the total cooling of the whole structure since
the line emission from optically-thin regions is far more efficient and
therefore determines the overall cooling budget.

\subsection{Low Energy-Density Environments}

The temperature profiles of an isolated cloud of gas at various constant
densities $n_{\rm H}$ and in a low $T_{\rm rad}$ environment are shown in
Fig. \ref{fig:profiles}$c$.
These profiles are calculated for a radiation field with a Compton
temperature $T_{\rm C} = 4 \times 10^{7}$ K and with a self-absorption
turnover frequency $\nu_{\rm t} = 3 \times 10^{13}$ Hz, which is the same
as that used in Fig. \ref{fig:profiles}$a$, but corresponds to a maximum
brightness temperature $T_{\nu_{\rm t}} = 9 \times 10^{8}$ K that is
considerably lower owing to the lower energy density.

Here, the gas structure is always optically-thin above temperatures of
$\mbox{a few } \times 10^{4}$ K.
The temperature of an optically-thin cloud at its illuminated face only
begins to approach the Compton limit when the density falls well below
$10^{14} \mbox{ cm}^{-3}$, with induced scattering contributing about
$25\%$ to the total Compton heating.
At higher densities, essentially all the cloud heating is due to free-free
absorption, although photoabsorption becomes important as the column density
of the gas increases.
The dominant cooling process at temperatures $< 10^{6}$ K is line emission
by collisionally-excited FeIX (171 \AA) and, to a lesser extent, by
collisionally-excited MgVII (433 \AA) and OIV (789 \AA).
These lines can contribute up to $50\%$ of the total cooling of the gas.
As in the case for the profiles in Fig. \ref{fig:profiles}$b$, HII and HeII
nuclei regulate the cooling as the temperature approaches $10^{5}$ K, the
main processes being bremsstrahlung emission, recombination (including
induced HII recombination), Ly $\alpha$ emission and HeII collisional
ionization.
As the cloud thickness increases and the temperature drops below $10^{5}$ K,
heavier nuclei regulate both the cooling, through collisional ionizations,
and the heating, through photoionization by soft X-rays.

\subsubsection{Summary}

In this Section, we have presented a detailed study of the temperature and
ionization structure of dense gas subjected to radiation that represents
a typical primary source expected in the central continuum-forming region
of AGN.
We have examined and compared three different cases which encompass the
high energy-density and the high brightness temperature regimes that are
relevant to a primary radiation field in these central environments.
The temperature profiles ($T_{\rm cld}$ vs. $R_{\rm cld}$) we have plotted
show that cool temperatures and low column densities are indeed compatible
thermal and radiative equilibrium properties of dense gas in the central
regions of AGN.
The dominant heating process is free-free absorption, while the cooling
is chiefly due to collisionally-excited line emission.

In high energy-density environments, we have found that additional heating
due to induced Compton scattering can make a significant difference to the
internal ionization structure of optically-thin gas at even the highest
densities, when the scattering competes with radiative (free-free)
absorption.
When the primary radiation field has a high brightness temperature,
induced scattering effectively raises the temperature of gas by as much
as two orders of magnitude at the illuminated edge of the structure.
The gas is then able to cool efficiently through a much wider range of
successively lower ionization stages of heavy nuclei.
In low energy-density environments, the temperature and corresponding
ionization levels of dense gas are below the potential of most metals,
so that the line cooling is mainly due to hydrogen and helium nuclei.
Photoionization heating is also more important in this case.

\section{The Thermal Stability of a Cloud}
\label{sec:stability}

The solutions $T_{\rm cld}$ to the electron temperature evaluated at each
point in a cloud of gas by balancing all the radiative heating and cooling
processes under isochoric conditions describe thermal equilibria which can
be either stable or unstable.
The gas is said to be in a thermally stable phase if the response to a
perturbation $\delta T > 0$ is to immediately cool down.
In other words, the criterion for thermal stability of a static gas is
that the temperature derivative of the net volume cooling rate (i.e.
total cooling $C$ minus total heating $H$) satisfies \cite{Field65}
\[
\frac{d (C-H)}{d T} > 0
\]
at a local equilibrium temperature and under some physical constraint,
such as the isochoric condition assumed here.
An unstable equilibrium corresponds to a physical state in which the
cloud gas is susceptible to perturbations that grow with time and
eventually break up the gas into multiple stable sub-phases, thereby
precluding the existence of clouds at that local equilibrium temperature.

It has been demonstrated that gas at temperatures above $10^{4}$ K can
exist in a thermally-stable state under the conditions expected in the
BLR \cite{McCray79,KMT81,GuilbertFabianMcCray83}.
Unlike this more distant gas, which is responsible for the optical
and UV lines that characterize the spectra of radio-quiet AGN,
photoionization and recombination do not regulate the overall thermal
and ionization balance of denser gas distributed throughout the
central source region. 
This is because at high densities and low column densities, free-free
absorption dominates photoabsorption and collisional rates are faster
than spontaneous recombination rates.
It is not obvious, therefore, whether thermal instability precludes the
existence of gas at the corresponding equilibrium temperatures which
are found under these conditions.

Clearly, in this situation, the physical state of dense gas cannot be
appropriately described using the standard photoionization parameter
$\xi = L/n_{e} r^{2}$ (where $L$ is the ionizing luminosity), which
assumes photoionization equilibrium.
Hence, the thermal stability of the gas cannot be examined by plotting
the corresponding radiative equilibrium curves ($\xi$ vs. $T_{\rm cld}$).
Instead, thermally unstable equilibria can be identified as solutions
$T_{\rm e}$ to the equation $C-H = 0$ at which the radiative cooling
curve ($C-H$ vs. $T_{\rm e}$) has a negative gradient, indicating that
the local heating rate rises faster with temperature than does the local
cooling rate.
In general, such cooling curves have complex shapes as they trace the
dominant processes for the corresponding temperature and ionization
state of the gas.

The isochoric cooling curves for dense gas irradiated by a primary
AGN source typically show thermally-stable equilibria over the range 
$\sim 10^{5}$ K to $\ltapprox 10^{6}$ K and at temperatures above
(a few)$\times 10^{6}$ K \cite{Kuncic96}.
It does not necessarily follow that these thermal and radiative equilibria
are also stable under isobaric conditions, since there may be two or more
combinations of ($n_{\rm H} , T_{\rm cld}$) that correspond to a given
pressure, so that a multiphase equilibrium is possible.
However, since the most plausible confinement mechanism for dense
clouds in the central continuum-forming region of AGN is via magnetic
stresses (CFR92), then the gas can maintain a range of possible thermal
pressures while the total (thermal + magnetic) cloud pressure balances
the total external pressure (which is likely to be magnetically-dominated
-- see \citeNP{KBR96}).
Thus, isobaric instability at a fixed thermal pressure does not preclude
the existence of the type of clouds in question here.
On the other hand, we note that magnetic confinement does not alleviate
isochoric instability.

The same may also hold for the more distant BLR clouds, where magnetic
fields may contribute a substantial fraction to the external pressure if
additional heating mechanisms in the hot, intercloud plasma are unable
to provide the large thermal pressures required to maintain pressure
equilibrium with the clouds \cite{MathewsFerland87}.
Typical cooling curves for the photoionized gas which comprises the
BLR clouds in this region indicate that equilibrium temperatures
$\sim 10^{5}$ K may be isobarically unstable
(see Figs. 2 and 3 in \citeN{KMT81}, for example).
However, if these clouds are magnetically-confined \cite{Rees87} and
can thereby maintain a range of possible pressures, then there is a
corresponding range of possible stable phases.
Indeed, this may even explain the wide range of reprocessing features
(e.g. warm absorber, broad absorption lines) that are inferred to
originate from roughly the same region ($\sim 0.1 - 1$ pc).

\section{Emergent Spectra from a Distribution of Clouds}

In the previous Sections, we have examined how very dense gas subjected
to an intense radiation field, representing a typical primary AGN source,
can cool to temperatures substantially lower than the Compton value
before becoming Thomson thick.
It is now of particular interest to predict the emergent spectra from
a source region containing such cool, dense substructure and to then
compare these spectra with current observations in an attempt to test the
hypothesis that these reprocessing clouds may indeed exist in the central
regions of AGN.

The simplest method is to combine the emission from a homogeneous
distribution of individual clouds into a composite spectrum which
can then be directly compared with observed spectra \cite{Netzer90}.
In addition to the parameters required to model an isolated cloud
($n_{\rm H}, R_{\rm cld}$) in a prescribed radiation field, a volume
filling factor $f$ must also be specified.
We have constructed a detailed model for such a cloud system; this
is described next, followed by a comparison of the predicted emergent
spectra with current observations.

\subsection{The Cloud Model}

Adopting the method of FR88, we assume that homogeneous, dense clouds
embedded within a plane-parallel source region of size $R$ absorb and
reprocess the primary radiation emitted by surrounding relativistic
particles.
We are particularly interested in the specific scenario of many
small-scale clouds which occupy only a tiny fraction $f$ of the
total volume, but which reprocess a significant fraction of 
the primary radiation.
Hence, we assume a covering factor of unity, so that there is at least one
cloud along every line of sight.
In this picture, the volume filling factor is equivalent to the line of
sight filling factor, with $f = {\cal N}_{\rm los} R_{\rm cld} / R$.
The total number of clouds along the line of sight, ${\cal N}_{\rm los}$,
is defined by specifying a total line of sight hydrogen column density
$N_{\rm tot} = {\cal N}_{\rm los} n_{\rm H} R_{\rm cld}$
(a quantity that can be observationally constrained).
The opacity of a cloud, $\kappa_{\nu}$, then implies a total
line of sight optical depth for the system,
$\tau_{\nu} = {\cal N}_{\rm los} \kappa_{\nu} R_{\rm cld}$.

\begin{figure*}
\hspace{0.01\textwidth}
\epsfysize=0.5\textwidth
\setbox2=\vbox{\epsfbox{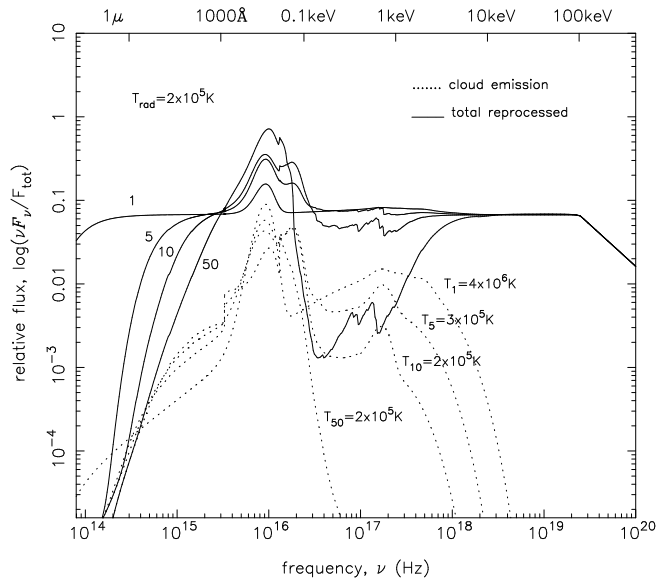}}
\rotr2
\caption{Emitted and total reprocessed (attenuated primary + cloud emission)
spectra from a typical cloud along lines of sight with 1, 5, 10 and 50
identical clouds.
The corresponding cloud temperature is indicated.} 
\label{fig:gridspec}
\end{figure*}

The emergent spectrum from the source region is calculated by integrating
the source function $S_{\nu}$ weighted by the optical depth, as prescribed
by the Schwarzschild-Milne solution to the plane-parallel radiative transfer
equation,
\[ 
F_{\nu} = 2 \int_{0}^{\tau_{\nu}} \! dt_{\nu} \,
    S_{\nu} (t_{\nu}) \, E_{2}(t_{\nu}) \, ,
\]
where $E_{2}(t) = e^{-t} - t E_{1}(t)$ is the standard exponential integral
function (of second-order) and where the first-order function
$E_{1}(t) \equiv \int_{t}^{\infty} dt' \, t'^{-1} e^{-t'}$
is calculated with a numerical integration.
Since the clouds are assumed to be the only source of opacity, the
source function for the region can be expressed as
\[ 
S_{\nu} = \frac{(1-f)\epsilon_{\nu}^{\rm prim}}{f \kappa_{\nu}}
        + S_{\nu}^{\rm cld} \, ,
\]
where $\epsilon_{\nu}^{\rm prim}$ is the (constant) volume emissivity of
the primary radiation and $S_{\nu}^{\rm cld}$ is the cloud source function,
which explicitly includes line emission (this was not taken into account by
FR88).
A gaussian profile is used for the lines, assuming the clouds are in random
motion with a bulk velocity $V \sim 0.2 c$, which is of the order of the
free-fall velocity at about 20 gravitational radii.

With homogeneous clouds, the emergent spectrum simplifies to
\[
F_{\nu} = \left( \frac{F_{\nu}^{\rm prim}}{2\tau_{\nu}}
      + S_{\nu}^{\rm cld} \right)
      \left[ 1 - 2E_{3}(\tau_{\nu}) \right] \, ,
\]
where $F_{\nu}^{\rm prim} = 2 (1-f) \epsilon_{\nu}^{\rm prim} R$ is the
(optically-thin) flux of the primary radiation field.

The attenuation described by the (third-order) exponential function
$E_{3}(\tau_{\nu})$ converges in the limit of large and small optical
depth according to
\[
1 - 2 E_{3} (\tau_{\nu}) \sim \left\{
  \begin{array}{ll}
        2 \tau_{\nu}  \, ,  &  \tau_{\nu} \ll 1  \\
        1             \, ,  &  \tau_{\nu} \gg 1 \, .
  \end{array}  \right.
\]
This behaviour implies corresponding regimes in the equation
for the emergent spectrum, 
\[
F_{\nu} \sim \left\{
  \begin{array}{lc}
        F_{\nu}^{\rm prim} + {\cal N}_{\rm los}F_{\nu}^{\rm cld} &
          \tau_{\nu} \ll 1 \\ 
        F_{\nu}^{\rm prim} / 2\tau_{\nu} +
          S_{\nu}^{\rm cld} & \tau_{\nu} \gg 1  \, ,
  \end{array}  \right.
\]
where
$F_{\nu}^{\rm cld}=S_{\nu}^{\rm cld} [1-2E_{3}(\kappa_{\nu}R_{\rm cld})]$
is the flux emitted by a typical cloud.
The source region is optically-thin to absorption ($\tau_{\nu} \ll 1$)
in the EUV band (as well as in the hard X-ray band, where the primary
radiation remains unattenuated).
The contribution by ${\cal N}_{\rm los}$ clouds to the emergent spectrum
is just the sum of their optically-thin spectra (chiefly due to lines and
some bremsstrahlung).
Since individual clouds are optically-thick at and below the FIR band,
the reprocessed spectrum due to many clouds always emerges as a complete
blackbody (in the Rayleigh-Jeans limit) at these frequencies
(i.e. $\tau_{\nu} \gg 1$).
At intermediate values of $\tau_{\nu}$, the emergent spectrum depends more
sensitively on the total number of clouds; the optical depth to absorption
at optical/UV frequencies and at soft X-ray energies increases with
${\cal N}_{\rm los}$.

A direct iteration of the cloud source function $S_{\nu}^{\rm cld}$ is
strictly required for self-consistentcy with the emergent spectrum when
${\cal N}_{\rm los} \gg 1$.
We approximate $S_{\nu}^{\rm cld}$ by taking into account that a typical
cloud does not `see' all of the primary radiation, owing to shielding by
other clouds, which generate a local reprocessed radiation field.
Assuming the filling factor and intercloud medium are uniform, the
fraction of primary radiation seen by a cloud is approximately
$d/R = 1/{\cal N}_{\rm los}$, where $d$ is the smallest region in which
there is one cloud along every line of sight.

Using {\small CLOUDY}, we construct a working grid model in which the
spectrum incident upon a cloud along a line of sight with a total of
${\cal N}_{\rm los}$ clouds is taken to be 
\[
F_{\nu}^{\rm inc} = \left( \frac{1}{{\cal N}_{\rm los}} \right)
    F_{\nu}^{\rm prim} +
                \left( 1 - \frac{1}{{\cal N}_{\rm los}} \right)
    F_{\nu}^{\rm repr}  \, ,
\]
where $F_{\nu}^{\rm repr}$ is the reprocessed flux (attenuated primary
plus emitted thermal) generated by ${\cal N}_{\rm los} - 1$ clouds.
We then take $S_{\nu}^{\rm cld}$ to be the source function of the cloud
irradiated with this incident spectrum.
The assumption of a covering factor of unity implies that the only primary
radiation `seen' directly is that which is transparent to radiative
absorption (this will be at hard X-ray energies).
However, it is conceivable that the region is not homogeneous and that
there could be some primary radiation emitted in an outer boundary
layer where clouds are too few in number to absorb the soft X-rays.
In any case, it is straightforward to take into account that a fraction of
the primary radiation may be generated in regions which do not intercept
any reprocessing material.

Our procedure gives a much more accurate treatment than that of FR88,
where the spectral properties (opacity and emissivity) of the reprocessing
gas were taken to be those calculated for an isolated cloud irradiated by
the primary source (i.e. without considering the local reprocessed field).
As a consequence of our treatment, the internal temperature gradient in
each cloud is significantly lower than that in the case of an isolated
cloud, as examined in Section 3 (and see in particular the temperature
profiles in Fig. \ref{fig:profiles}$a$).
Similarly, the spectral properties and emergent flux from successive
clouds converge towards local thermodynamic equilibrium as more of the
radiation is reprocessed.

In Fig. \ref{fig:gridspec}, we plot the spectra emerging from lines of
sight with different total numbers of optically-thin clouds that are
irradiated by an incident spectrum $F_{\nu}^{\rm inc}$ as defined
in this grid model.
The clouds are identical, with $n_{\rm cld} = 10^{18}{\rm cm}^{-3}$ and
$R_{\rm cld} = 1 \times 10^{3}$ cm, and the primary radiation field is
the same as that assumed in Fig. \ref{fig:profiles}$a$, for which isolated
clouds have the largest internal temperature gradient.
The dotted curves are the spectra emitted by a cloud along lines of sight
with ${\cal N}_{\rm los} = 1,5,10,50$.
The solid curves represent the corresponding total reprocessed spectra
obtained by summing the attenuated primary flux with the emitted flux.

This plot shows how thermal coupling with the local reprocessed
radiation field generated when ${\cal N}_{\rm los} \gg 1$ gives rise to
much cooler clouds than when ${\cal N}_{\rm los} = 1$.
Because of the much higher values of $T_{\rm cld}$, the spectrum emitted
by an isolated, optically-thin cloud along a line of sight is chiefly due
to bremsstrahlung and line radiation at soft X-ray and EUV energies.
As ${\cal N}_{\rm los}$ increases and $T_{\rm cld}$ converges toward
$T_{\rm rad}$, however, the emission spectrum of a typical cloud along
the line of sight becomes increasingly less important at soft X-ray
energies and approaches a blackbody at low frequencies.
The emission is in the UV, where the Ly edge emerges, and in the EUV,
where the heavy element lines emerge.
These are the only spectral `windows' from which the reprocessed
radiation can emerge, since thermal self-absorption prevails at lower
frequencies where the optical depth to free-free absorption is $\gg 1$,
while photoabsorption  depletes the primary soft X-rays, which cannot be
replenished by thermal emission owing to the decrease in $T_{\rm cld}$.
The hard X-rays remain unattenuated since there is no Compton scattering.

\begin{figure*}
\hspace{0.01\textwidth}
\epsfysize=0.8\textwidth
\setbox3=\vbox{\epsfbox{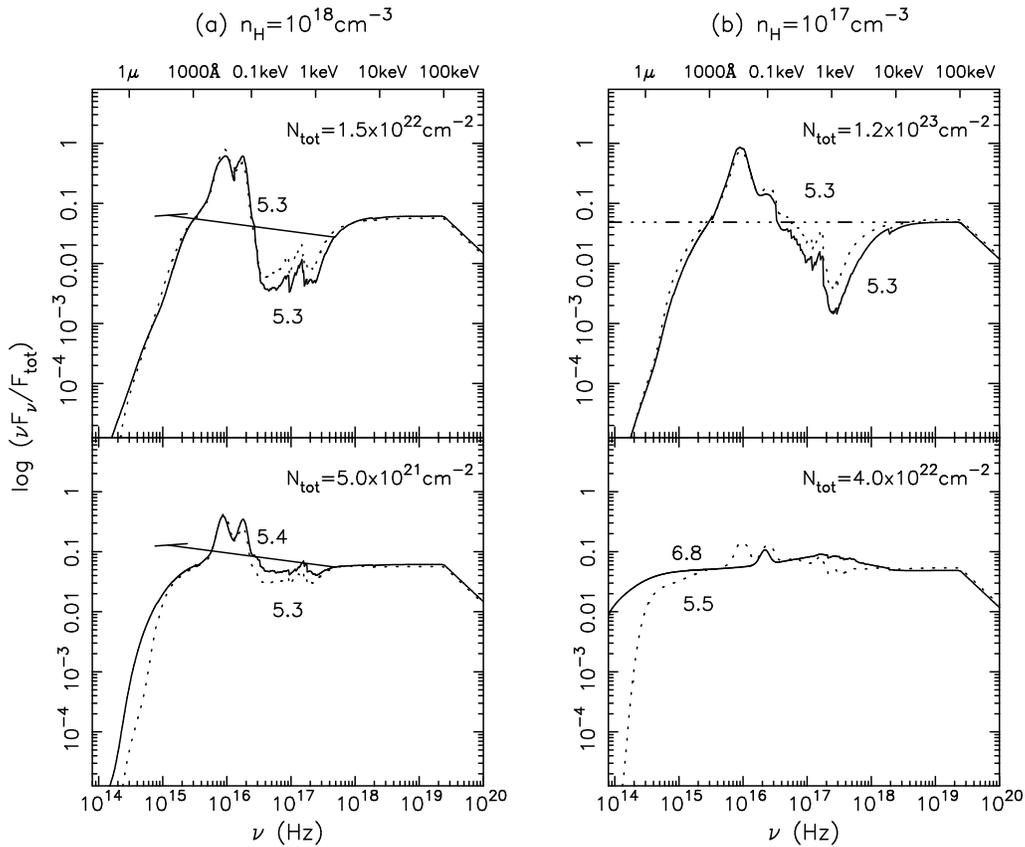}}
\rotr3
\caption{Emergent spectra from a source region in which primary radiation
with an energy-density temperature $T_{\rm rad} = 2 \times 10^{5}$ K and
with a maximum brightness temperature $T_{\nu_{\rm t}} = 2 \times 10^{11}$ K
is reprocessed by a total line-of-sight column density $N_{\rm tot}$ of
homogeneous clouds with density $n_{\rm H}$ and size $R_{\rm cld}$, where 
(a) $R_{\rm cld}=1 \times 10^{3}\,{\rm cm}\, , 5 \times 10^{3}\,{\rm cm}$;
and
(b) $R_{\rm cld}=2 \times 10^{5}\,{\rm cm}\, , 4 \times 10^{5}\,{\rm cm}$
(the total line-of-sight number of clouds is
${\cal N}_{\rm los} = N_{\rm tot} / n_{\rm H} R_{\rm cld}$).
Small and large $R_{\rm cld}$ are distinguished by solid and dotted curves,
respectively, and the logarithm of the resulting cloud temperature,
$\log T_{\rm cld}$ is indicated.
Also shown is a hypothetical power-law indicating the relative fluxes of
the optical/UV and X-ray bands using a spectral index $\alpha_{\rm ox}=1.14$
obtained from recent observations.
The average observed optical/UV slope is also indicated.
The dash-dots line represents the primary spectrum.}
\label{fig:spectra1}
\end{figure*}

\subsection{Results}

Plots of the predicted emergent spectra are presented in
Figs. \ref{fig:spectra1}, \ref{fig:spectra2} and \ref{fig:spectra3}.
The primary spectrum defined in Section 2 is shown for reference in
each figure (dash-dot line) and has the same
$T_{\rm rad} = 2 \times 10^{5}$ K in Figs. \ref{fig:spectra1} and
\ref{fig:spectra2}, but with
$\nu_{\rm t} = 3 \times 10^{13}$ Hz and $\nu_{\rm t} = 5 \times 10^{14}$ Hz,
respectively, while Fig. \ref{fig:spectra3} has
$T_{\rm rad} = 5 \times 10^{4}$ K, with $\nu_{\rm t} = 3 \times 10^{13}$ Hz.
The spectra are plotted as flux per logarithmic frequency interval,
$\nu F_{\nu}$, relative to the total integrated flux,
$F_{\rm tot} = acT_{\rm rad}^{4}$, thus giving the best indication of
the frequency ranges where most of the radiation is absorbed and reemitted
by the clouds.
Also shown in the figures is a hypothetical power-law from the optical
(2500 \AA) to X-ray (2 keV) with a spectral index $\alpha_{\rm ox}$
that is taken from current observations \cite{Puch96}, as well as the 
slope of the observed optical spectra (taken from the same data), where
the blue bump  component is strongest.
In Fig. \ref{fig:spectra3}$b$, the optical to X-ray power-law is drawn
relative to 2 keV X-rays in the primary spectrum, owing to the lack of
flux in the reprocessed spectrum.
This power-law therefore indicates the maximum relative flux in the
optical/UV in this spectrum.

The spectra shown in each plot are calculated for a constant value of
$n_{\rm H}$.
The solid and dotted curves in Figs. \ref{fig:spectra1} and
\ref{fig:spectra2} distinguish between small and large values
of $R_{\rm cld}$, respectively, that correspond to clouds which
are optically-thin and optically-thick to free-free absorption
at $10^{15}$ Hz, as examined in Section 3.
The lower panels in these figures show spectra for lower values of
$N_{\rm tot}$ that correspond to a single optically-thick cloud
(${\cal N}_{\rm los}=1$), so the emergent spectra closely resemble that
expected for a `slab' (since the covering factor is implicitly assumed to
be unity) and can be directly compared with the spectra calculated for
smaller $R_{\rm cld}$ and higher ${\cal N}_{\rm los}$.
Fig. \ref{fig:spectra3}$a$ also shows the integrated spectrum due to a
single cloud along the line of sight, while Fig. \ref{fig:spectra3}$b$
shows the corresponding spectrum for ${\cal N}_{\rm los} = 100$.
The parameters used in Fig. \ref{fig:spectra3}$c$ are specifically
chosen to represent the Thomson-thin, free-free emission model proposed by
\citeN{Barvainis93} for the optical/UV blue bump component in AGN spectra.

\subsubsection{Interpretation}

\begin{figure*}
\hspace{0.01\textwidth}
\epsfysize=0.8\textwidth
\setbox4=\vbox{\epsfbox{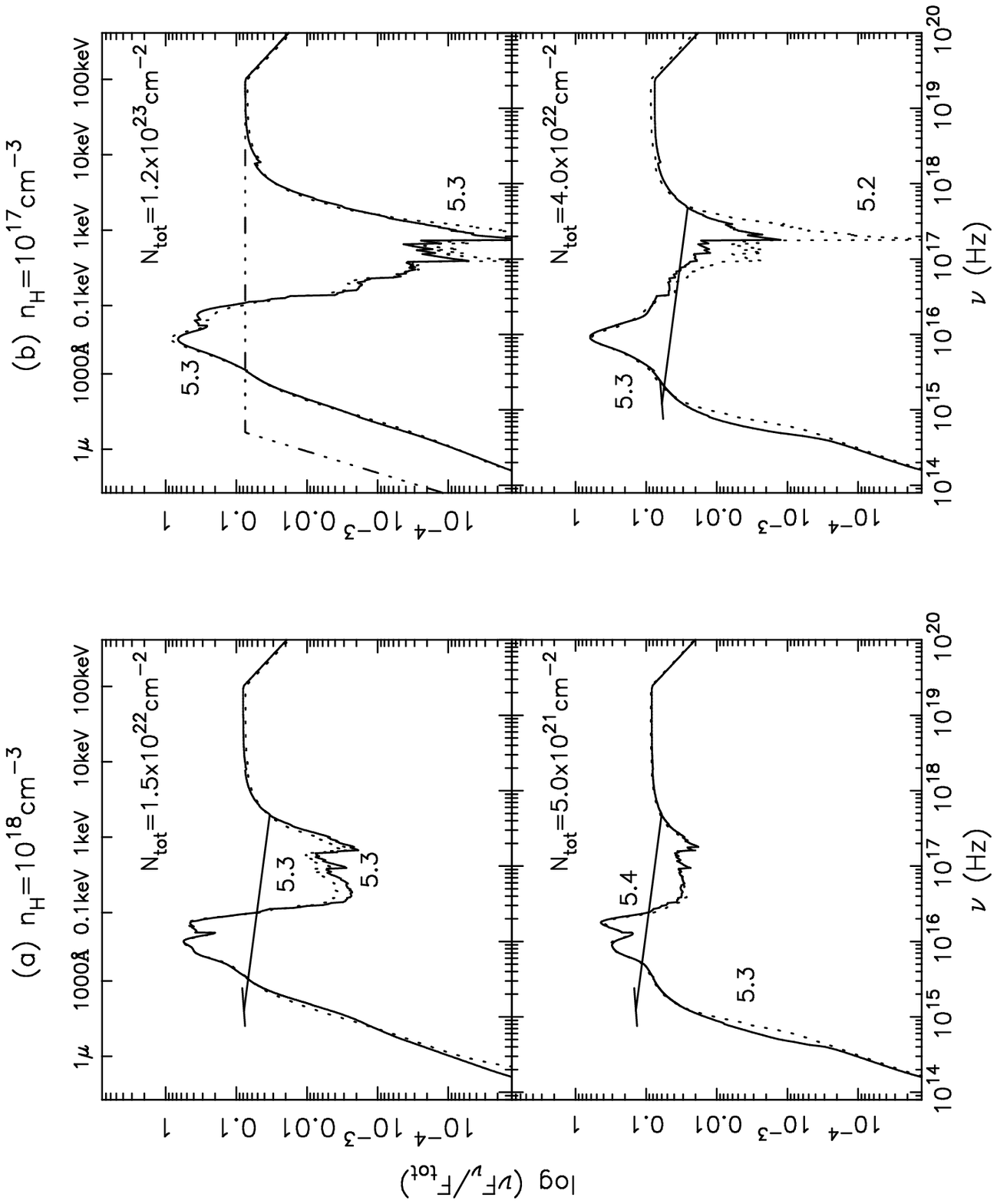}}
\rotr4
\caption{Emergent spectra from a source region in which primary radiation
with an energy-density temperature $T_{\rm rad} = 2 \times 10^{5}$ K and
with a maximum brightness temperature $T_{\nu_{\rm t}} = 6 \times 10^{7}$ K
is reprocessed by a total line-of-sight column density $N_{\rm tot}$ of
homogeneous clouds with density $n_{\rm H}$ and size $R_{\rm cld}$, where 
(a) $R_{\rm cld}=1 \times 10^{3}\,{\rm cm}\, , 5 \times 10^{3}\,{\rm cm}$;
and
(b) $R_{\rm cld}=2 \times 10^{5}\,{\rm cm}\, , 4 \times 10^{5}\,{\rm cm}$
(the total line-of-sight number of clouds is
${\cal N}_{\rm los} = N_{\rm tot} / n_{\rm H} R_{\rm cld}$).
Small and large $R_{\rm cld}$ are distinguished by solid and dotted curves,
respectively, and the logarithm of the resulting cloud temperature,
$\log T_{\rm cld}$ is indicated.
Also shown is a hypothetical power-law indicating the relative fluxes of
the optical/UV and X-ray bands using a spectral index $\alpha_{\rm ox}=1.14$
obtained from recent observations.
The average observed optical/UV slope is also indicated.
The dash-dots line represents the primary spectrum.}
\label{fig:spectra2}
\end{figure*}

\begin{figure*}
\begin{minipage}{1.0\textwidth}
\epsfysize=1.0\textwidth
\setbox5=\vbox{\epsfbox{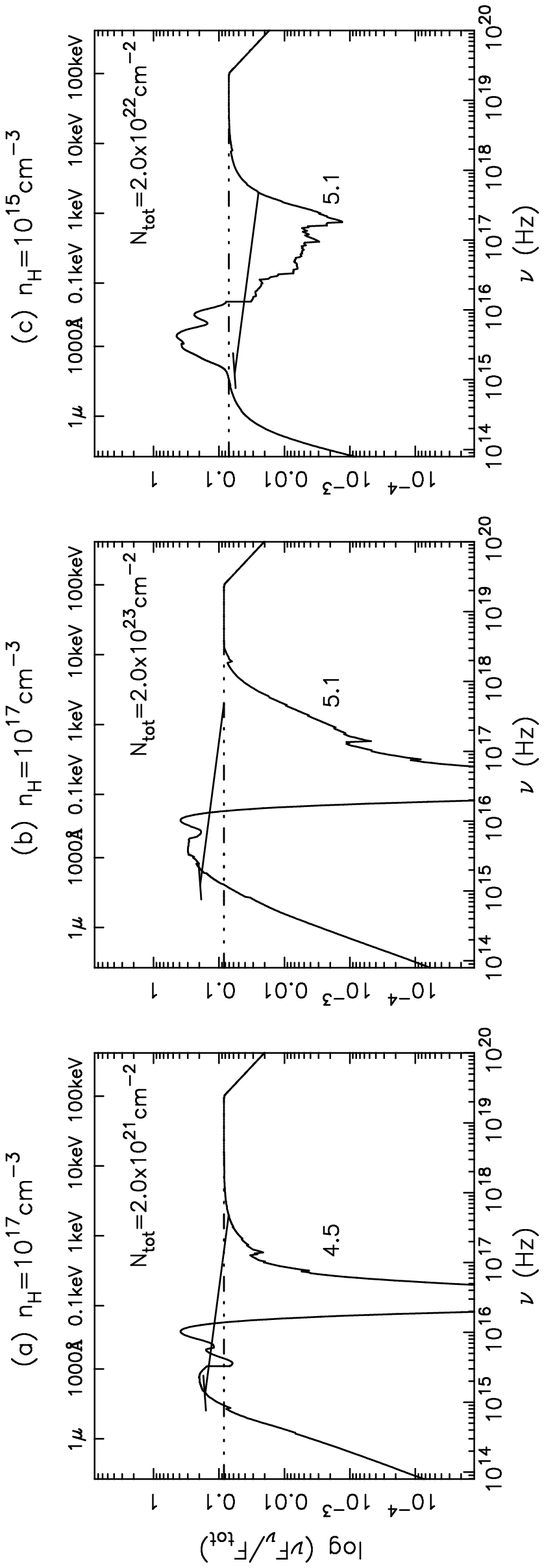}}
\rotr5
\caption{Emergent spectra from a source region in which primary radiation
with an energy-density temperature $T_{\rm rad} = 5 \times 10^{4}$ K and
with a maximum brightness temperature $T_{\nu_{\rm t}} = 9 \times 10^{8}$ K
is reprocessed by a total line-of-sight column density $N_{\rm tot}$ of
homogeneous clouds with density $n_{\rm H}$ and size $R_{\rm cld}$, where \,
(a) $R_{\rm cld}=2 \times 10^{4}\,{\rm  cm}$;
(b) $R_{\rm cld}=2 \times 10^{4}\,{\rm cm}$;
and
(c) $R_{\rm cld}=2 \times 10^{7}\,{\rm cm}$
(the total line-of-sight number of clouds is
${\cal N}_{\rm los} = N_{\rm tot} / n_{\rm H} R_{\rm cld}$).
The logarithm of the resulting cloud temperature, $\log T_{\rm cld}$ is
indicated.
Also shown is a hypothetical power-law indicating the relative fluxes of
the optical/UV and X-ray bands using a spectral index $\alpha_{\rm ox}=1.14$
obtained from recent observations.
The average observed optical/UV slope is also indicated.
The dash-dots line represents the primary spectrum.}
\label{fig:spectra3}
\end{minipage}
\end{figure*}

The predicted emergent spectra clearly exhibit the common property of an
energetically-significant reprocessed component which results from the
direct reradiation of energy that has been depleted from the primary
radiation field in the optical/UV band and below (by free-free absorption
and in some cases, induced Compton scattering) and also in the soft X-ray
band (by photoabsorption).
A Rayleigh-Jeans component ($\propto \nu^{2}$) emerges when free-free
absorption has appreciably depleted the low-frequency primary radiation
(see the upper panels in Figs. \ref{fig:spectra1} and \ref{fig:spectra2}
and also Figs. \ref{fig:spectra3}$a$ and \ref{fig:spectra3}$b$).

Photoabsorption of soft X-rays becomes the chief heating mechanism for
dense gas once the low-frequency radiation is efficiently removed.
This is demonstrated by comparing the spectra in Fig. \ref{fig:spectra1}$b$
with those in Fig. \ref{fig:spectra1}$a$, which are calculated for lower
$n_{\rm H}$ and higher values of $N_{\rm tot}$, but which exhibit less
absorption of the primary soft X-rays because not all of the low-frequency
photons have been removed.

The difference in gas temperatures is also an important factor determining
the relative importance of free-free absorption and photoabsorption.
Below $\sim 10^{5}$ K, when the ionization state is lower and hence, there
are fewer free electrons, photoabsorption can be stronger than free-free
absorption.
This is evident in Fig. \ref{fig:spectra3}.
Similarly, the soft X-ray extinction also becomes appreciable when the
primary radiation field has less low-frequency energy available for
free-free absorption.
This is demonstrated by comparing Fig. \ref{fig:spectra1} with
Fig. \ref{fig:spectra2}, which is for a higher FIR turnover frequency,
$\nu_{\rm t}$, and which shows how less energy in the primary spectrum
available for low-frequency heating is compensated for by more soft X-ray
photoelectric heating.

The photoabsorption spectral features which emerge most notably are
L-edges of C and O ($\sim 0.4$ keV) and in some cases, an Fe K-edge
in the X-ray band ($\sim 0.8$ keV) and a Ly edge due to collisional
ionization of HeII (228 \AA).
Probably the most striking result shown by these spectra, however, is
that for all feasible cloud parameters, line radiation is significant
and in most cases dominates the continuum radiation.
The bulk of the predicted lines emerge in the EUV band, which is the
only `window' in the broadband spectra that is unaffected by radiative
absorption due to the clouds.
All the lines are collisionally-excited heavy element lines (spontaneous
recombination radiation is negligible by comparison, as discussed earlier
in Section 4), with the most energetically-significant lines typically
being FeIX (171 \AA), HeII (304 \AA) and MgVII (433 \AA).

In the high $T_{\rm rad}$ case (Figs. \ref{fig:spectra1} and
\ref{fig:spectra2}), almost all the lines are resonance lines of Fe
in various ionization states and the total line emission dominates the
continuum emission both when there are many optically-thin clouds and when
there are fewer optically-thick clouds at the same total column density.
This is because in such an intense radiation field, line emission is the
most efficient means by which the clouds can cool from temperatures which
are initially very high.
The case when optically-thin clouds are insufficient in number to reprocess
an energetically-significant amount of radiation and thereby cool appreciably
is illustrated in Fig. \ref{fig:spectra1}$b$, with
$N_{\rm tot}= 4 \times 10^{22} \,{\rm cm}^{-2}$ (solid curve).
The temperature of the optically-thin clouds is so hot ($\sim 10^{6.8}$ K)
that a hump emerges at X-ray energies due to bremsstrahlung emission and
indeed, this exceeds the line emission.
The spectrum of the thicker clouds (dotted curve) shows the lines
responsible for their much cooler temperature ($\sim 10^{5.5}$ K).

In the low $T_{\rm rad}$ case shown in Fig. \ref{fig:spectra3}, fewer Fe
lines are emitted owing to the lack of high-ionization state Fe ions.
The collisionally-excited HeII (304 \AA) line, on the other hand, is a
prominent feature as is the HeII Lyman (228 \AA) edge due to collisional
ionization.
A H Ly edge is also visible either in absorption or, at higher $N_{\rm tot}$,
in emission (compare Figs. \ref{fig:spectra3}$a$ and \ref{fig:spectra3}$b$).
This edge emerges because the cloud temperature is below $10^{5}$ K and
free-free absorption is incapable of thermalizing the spectrum at optical
frequencies when $N_{\rm tot}$ is low.

\subsubsection{Testing the Free-Free Emission Model}

Fig. \ref{fig:spectra3}$c$ illustrates a specific case using parameters
representative of an optically-thin thermal model proposed by
\citeN{Barvainis93}.
In this model, the optical/UV blue bump component is attributed to thermal
bremsstrahlung emission by a source region that is occupied by clouds with
${\cal N}_{\rm los}\sim 1$ and that is optically-thin to free-free
absorption above optical frequencies.
The main shortcoming of this model is that it is based on an analytical
treatment in which free-free emission is assumed to be solely responsible
for the blue bump.
The cloud parameters (temperature and size) are therefore not calculated
under thermal and radiative equilibrium conditions and hence, are not
self-consistent with the ionization balance of the gas and the ambient
reprocessed radiation field.
By taking this into account, our results demonstrate that free-free
radiation is not the dominant component of the emission from the gas,
owing to significant contributions by bound-free and bound-bound (line)
emission.
For the representative parameters of this model, we find that
photoionization contributes up to $35\%$ of the total heating budget
(this is evident in the predicted spectrum in Fig. \ref{fig:spectra3}$c$),
while the total cooling budget is dominated by line emission, which
accounts for up to 3 times more than the continuum emission.
We find that this line-to-continuum ratio is even higher when the clouds
are thinner and ${\cal N}_{\rm los}\gg 1$.
Similar results demonstrating the inadequacies of free-free models for the
blue bump have also been obtained by \citeN{Collin96}.

\subsubsection{Comparison with Observations}

The predicted emergent spectra in Figs. \ref{fig:spectra1},
\ref{fig:spectra2} and \ref{fig:spectra3} indicate that reprocessing by
very dense, small-scale clouds is energetically-significant under a wide
range of conditions.
However, when compared with observed AGN spectra, such as those presented
in the atlas of SEDs by \citeN{Elvis94}, the predicted spectra show an
obvious lack of flux over a narrow, but crucial range of optical/UV
energies below $10$ eV (above $\sim 1200$ \AA), where the blue bump
component reaches a maximum and appears to roll over.
The same problem can be seen with the spectra predicted by the model of
FR88 (see their Fig. 8).

The discrepancy can be described quantitatively with the parameter
$\alpha_{\rm ox}$, which is usually defined as the slope of a hypothetical
power-law ($F_{\nu} \sim \nu^{-\alpha}$) extending over 2500 \AA - 2 keV
\cite{Tananbaum79}, giving a measure of the relative optical to X-ray flux
and thus, a measure of the strength of the blue bump component.
This power-law is drawn against some of the predicted spectra for comparison.
The spectral index used is $\alpha_{\rm ox} = 1.14$, which is the mean value
measured from the Rosat International X-ray/Optical Survey (RIXOS) of AGN
\cite{Puch96}.
Note, however, that we have not included any reprocessed component in X-rays,
which could flatten the 2-40 keV spectrum.
Also shown is the average slope measured in the optical band, with
$\alpha_{\rm opt} \sim 0.92$ from the same sample.
The inconsistency between our predicted spectra and the observed optical/UV
spectra is even more apparent when considering that the RIXOS sample of AGN
is considerably larger and less selective than other samples, which
typically show steeper optical to X-ray ratios, with an average
$\alpha_{\rm ox} \sim 1.4$
(e.g. the optically-selected PG quasars analysed by \citeNP{Elvis94}).

In the high $T_{\rm rad}$ case (Figs. \ref{fig:spectra1} and
\ref{fig:spectra2}), the emergent spectra are typically flux-limited
in the optical/UV due to the large line of sight optical depths to
free-free absorption at these frequencies.
The blackbody spectra clearly rise too steeply and lack the flux
required to account for the peak in the blue bump.
This deficiency is apparent also when the spectra are optically-thin
at these frequencies (see Fig. \ref{fig:spectra1}$b$, lower panel), since
free-free absorption then reprocesses very little of the primary radiation
and photoabsorption reprocesses even less, so that very little flux is
reemitted in the optical/UV band.

The closest agreement with the observed optical/UV spectra is found in
the low $T_{\rm rad}$ case shown by Fig. \ref{fig:spectra3}$a$.
However, the conditions in this case are quite different from those
which initially motivated our investigation.
Free-free absorption is less efficient (since there are fewer free
electrons), so that most of the primary radiation is reprocessed by
photoabsorption, with the soft X-rays reemitted as optical/UV radiation.
At higher $N_{\rm tot}$, atomic absorption depletes the X-rays towards
harder energies and the spectrum approaches a complete blackbody, as can
be seen in Fig. \ref{fig:spectra3}$b$.
This spectrum shows that even if some of the primary X-rays are replenished
in an outer boundary layer of the source region (where there are essentially
no clouds in these conditions), the ratio of the optical/UV to X-ray flux is
still inconsistent with the observations.
The spectrum rise too steeply at the frequencies where the blue bump
typically peaks.

It thus appears that consistency with the blue bump requires a total
line of sight optical depth to free-free absorption that is not large
at optical/UV energies, so that the flux is not blackbody-limited.
At the same time, some photoabsorption is required to produce an
energetically-significant reprocessed component in the optical/UV.

The only band that remains optically-thin to radiative absorption (apart
from hard X-rays) is the EUV, where the bulk of the radiation reprocessed
by the clouds emerges as extremely broad lines.
Although most extragalactic objects are virtually impossible to detect
at EUV energies because of strong galactic absorption by interstellar
gas, recent observations have been successful in identifying many AGN
in directions of low galactic neutral hydrogen as EUV sources
(see \citeNP{Fruscione95} and references cited therein).

For some objects, such as the radio-quiet quasars Ton S 180 and H1821+643,
the observed spectra at either end of the EUV clearly exhibit both a blue
bump that turns over at $\ltapprox 10$ eV and a steep soft X-ray excess at
$\ltapprox 0.5$ keV and the indication is that the optical/UV to soft X-ray
must be comprised of two or more components \cite{Wisotzki95,Kolman93}.
To date, however, very few EUV and soft X-ray observations have been
undertaken simultaneously and so the spectral slope into the EUV 
and its connection with the lower and higher energy bands remains
poorly known.
Such observations of the Seyfert galaxy Mrk 841 have revealed a soft X-ray
spectrum that extrapolates with a power-law fit well into the EUV.
This adds further support to the implication from non-simultaneous
observations that the bulk of the spectral power peaks in the EUV band
\cite{Nandra95} and that it is connected to the soft X-rays, possibly
forming a single `big bump' component that extends over to the optical/UV,
where it connects with the blue bump \cite{WalterFink93}.
Very recent results from RIXOS are also consistent with this idea
\cite{Puch96}.

Observations of the Seyfert 1 galaxy NGC 5548 show possible broad,
weak EUV lines that are difficult to explain in the framework of
photoionization models (\citeNP{Kaastra95} -- but see also
\citeNP{Marshall96}).
Consequently, it has been suggested that they are instead due to a dense,
collisionally-ionized warm absorber, with the emitting material then most
likely being in the form of many small clouds.
This fits in with the idea of a `very broad line region' (VBLR) containing
highly-ionized material, as proposed by \citeN{FKP90} in order to explain
similar broad-base lines observed in other AGN.
The line widths place the emitting region in the inner BLR.
In a global picture, it is plausible that the entire nuclear region of
radio-quiet AGN contains various forms of thermal reprocessing material
with radiative properties determined by the prevalence of photoionization
equilibrium in the BLR, while collisional processes become increasingly
more important in the denser, more central regions.

\subsection{Summary and Discussion}

We have presented a detailed analysis of very dense, geometrically-thin 
clouds that may exist in the central continuum-forming region of AGN and
that reprocess the primary radiation generated therein.
We have employed numerical methods to examine their physical
properties under thermal and radiative equilibrium conditions.
It was found that very dense gas can readily achieve cool temperatures
before reaching large column densities, which is consistent with the idea
of filaments or sheets confined by an external magnetospheric environment.
The heating of the gas is dominated by free-free absorption, although
photoabsorption can be important depending on the amount of primary
radiation available at low-frequencies and also on the amount of available
free electrons.

We have found that the total cooling budget is dominated by
collisionally-excited line radiation, essentially all of which is
emitted in the EUV band, which remains transparent to radiative
absorption at the column densities we considered.
We have also discussed the thermal stability of the gas in relation
to that of other reprocessing gas residing in more distant regions.
While there may be some thermal and radiative equilibria which are
unstable under isochoric or isobaric conditions, magnetic confinement
allows a much wider range of possible pressures, so that there is
always likely to be some stable equilibria.

We have calculated the emergent spectra from a source region in which
the primary continuum radiation is generated and in which dense clouds
are distributed homogeneously.
These spectra show that along lines of sight which are Thomson-thin,
dense clouds are capable of reprocessing an energetically-significant
amount of the primary radiation generated in the central continuum-forming
region of AGN.
However, our predicted spectra show that when dominated by free-free
absorption, this reprocessing cannot solely be responsible for the
observed optical/UV emission that is identified as the blue bump.

When the total line of sight optical depth to free-free absorption
is large at the crucial optical/UV frequencies where the blue bump
peaks, the emergent flux is blackbody-limited.
This is particularly the case for high energy-density environments,
since the large supply of free electrons ensures that reprocessing
by very dense, Thomson-thin gas is dominated by free-free absorption.
In low energy-density environments, the reprocessing is largely due to
photoabsorption and the optical/UV emission more closely resembles the
blue bump provided again that the spectrum is not completely blackbody
at these crucial frequencies.
This is consistent with recent results that suggest reprocessing by
Thomson-thick clouds can produce spectra which can account for the blue
bump as well as soft X-ray excesses and an X-ray reflection feature
\cite{Collin96}.
Because these Thomson-thick clouds are less dense, free-free absorption
is not important above IR frequencies and the optical/UV spectrum where
the blue bump peaks is entirely dominated by free-bound emission, rather
than being blackbody-limited as is the case here.
In the presence of strong photoabsorption, however, a covering factor of
unity requires the primary radiation to be generated in an outer layer of
the source region that is essentially free of this type of clouds, so that
some of the primary soft X-rays are detected
(but see also \citeNP{Collin96}).
A few hotter, thinner clouds in this outer layer would also contribute
to the observed soft X-ray emission.

Realistically, reprocessing clouds at the centres of AGN are probably
distributed inhomogeneously, with a range of densities and temperatures,
so that different lines of sight have different column densities and
different optical depths to radiative absorption and Thomson scattering.
Indeed, there is growing evidence that the BLR contains such a mixture
of optically-thin and optically-thick cloud structure, contrary to
long-standing ideas \cite{ShFerPet95}.

Despite the inability of very dense, Thomson-thin clouds to account for
the blue bump, the reprocessed spectra can still provide a significant
contribution to the SED of radio-quiet AGN.
The bulk of the radiation that is reprocessed by the clouds examined
here emerges as very broad EUV lines, which form the most prominent
and energetically-dominant spectral feature.
This result is particularly interesting in light of the current
observations which are now for the first time identifying radio-quiet
quasars and Seyferts in the largely inaccessible EUV band.
These observations provide convincing evidence for a link between the
optical/UV blue bump and soft X-ray excess via an extended `big bump'
which peaks in the EUV and which may be comprised of at least two
components.
Current detectors may soon further resolve this critical band to reveal
additional components and the results presented here would then provide
a test for the presence of thermal reprocessing structure in the form of
dense, Thomson-thin clouds in the innermost central regions of these
objects.

\section*{Acknowledgments}

We wish to thank Gary Ferland for his assistance with the workings of
his code {\small CLOUDY} and the referee, S. Collin-Souffrin, for helpful
comments on the manuscript.
We are grateful for financial support by the Cambridge Commonwealth Trust
(ZK), the Royal Society (AC \& MR) and the Italian {\small MURST} (AC).

\bibliography{mnrasmnemonic}
\bibliographystyle{mnrasv2}

\end{document}